\documentclass[compsoc,conference,a4paper,10pt,times]{IEEEtran}
\IEEEoverridecommandlockouts
\UseRawInputEncoding
 
\usepackage{array}
\usepackage{cite}
\usepackage{amsmath,amssymb,amsfonts}
\usepackage{algorithmic}
\usepackage{graphicx}
\usepackage{textcomp}
\usepackage{bmpsize}
\usepackage{xcolor}
\usepackage{xurl}
\usepackage[colorlinks=true,urlcolor=black]{hyperref}
\def\BibTeX{{\rm B\kern-.05em{\sc i\kern-.025em b}\kern-.08em
    T\kern-.1667em\lower.7ex\hbox{E}\kern-.125emX}}

\begin{document}

\title{Quantum Disruption: An SOK of How Post-Quantum Attackers Reshape Blockchain Security and Performance}

\author{
\IEEEauthorblockN{Tushin Mallick\textsuperscript{*}}
\IEEEauthorblockA{
Northeastern University\\
mallick.tu@northeastern.edu
}
\and
\IEEEauthorblockN{Maya Zeldin\textsuperscript{*}}
\IEEEauthorblockA{
Northeastern University\\
zeldinmaya@gmail.com
}
\and
\IEEEauthorblockN{Murat Cenk}
\IEEEauthorblockA{
Ripple Inc\\
mcenk@ripple.com
}
\and
\IEEEauthorblockN{Cristina Nita-Rotaru}
\IEEEauthorblockA{
Northeastern University\\
c.nitarotaru@northeastern.edu
}
\thanks{\textsuperscript{*}Equal Contribution.}
}

\maketitle

\begin{abstract}
As quantum computing advances, classical cryptographic systems---signature 
schemes, key exchange protocols, public-key encryption, and certain hash 
constructions---that secure most blockchain platforms come under threat, raising 
serious concerns about their long-term security and integrity. Transitioning 
to post-quantum primitives is rarely straightforward: their larger sizes and 
higher computation costs can have unintended consequences, and in some cases 
make such a transition impractical.

We examine the impact of post-quantum migration on blockchain systems across three dimensions. First, we dissect the cryptographic primitives most vulnerable to quantum attacks within the consensus, identity, and transaction-validation layers. Next, we survey proposed post-quantum adaptations and evaluate their practical feasibility. Finally, we assess how substituting classical primitives with post-quantum alternatives affects system performance, protocol behavior, and the underlying incentive and trust structures. Our findings show that 
PQ adoption is far from a drop-in replacement: it demands careful architectural 
reconsideration, as naive substitution risks destabilizing core protocol 
operations and weakening the security and efficiency guarantees blockchains 
rely upon.

\end{abstract}

\section{Introduction}

The emergence of quantum computing poses a critical challenge to classical 
cryptographic systems. Recent milestones---Microsoft's Majorana 1 quantum 
processor~\cite{microsoft2025majorana1} and Google's Willow chip~\cite{google2024willow}---suggest 
that practical quantum computation may soon be achievable. The risk is especially 
pronounced in blockchain systems, which depend on cryptographic primitives 
directly susceptible to quantum attacks, and which still secure hundreds of 
billions of dollars: as of mid-2026, DeFi platforms held roughly USD 100 
billion in total value locked~\cite{defillama2025tvl}, and Ethereum alone had a 
market capitalization of around USD 200 billion~\cite{coinmarketcap2025eth}. This 
concentration of value highlights the urgency of evaluating blockchain resilience 
against the quantum threat.

This value lives in a layered ecosystem of interacting entities. At its base are
\emph{Layer-1 (L1)} ledgers---independent networks such as Bitcoin and Ethereum
that run their own consensus, are secured by permissionless validators or miners,
and provide final settlement; above them, \emph{Layer-2 (L2)} rollups and payment
channels scale activity while inheriting L1 security, and \emph{cross-chain
bridges} connect otherwise isolated ledgers. The activity is concentrated in DeFi:
on-chain smart-contract protocols---decentralised exchanges (DEXs), lending
markets, derivatives venues, stablecoin issuers, and liquid-staking and
tokenisation platforms---that interoperate with off-chain entities such as wallets,
centralised exchanges, custodians, and price oracles, and are governed in part by
decentralised autonomous organisations (DAOs).

Shor's algorithm~\cite{shor1997algorithm} breaks the conventional signature 
schemes (RSA, DSA, ECDSA) used by every major chain. A few platforms have begun 
deploying quantum-resilient primitives in production: Algorand uses lattice-based 
Falcon signatures for its State Proofs, providing PQ-secure cross-chain 
attestations~\cite{algorand_stateproofs_falcon}, and Solana offers an opt-in 
Winternitz Vault based on the Winternitz One-Time Signature 
scheme~\cite{solana_quantum_resistant_vault_2025}. The transition to 
post-quantum cryptography (PQC) at protocol level, however, is non-trivial: 
cryptographic primitives are tightly integrated into identity management, 
transaction validation, and consensus, and replacing them creates cascading 
effects on interoperability, performance, and system stability. PQ algorithms 
also vary significantly in signature size, computational cost, and 
implementation maturity.

In this work, we present a comprehensive systematization of knowledge of the challenges faced by blockchains with respect to PQ preparedness. We consider seven L1 ledgers: Bitcoin \cite{bitcoin2025}, Ethereum \cite{ethereum2025}, Algorand \cite{algorand2025}, Monero \cite{monero2025}, Avalanche \cite{avalanche2025}, Solana \cite{solana2025}, and XRP Ledger (XRPL) \cite{xrpl2025}. These platforms were selected based on their prominence within the contemporary blockchain ecosystem, as reflected by market capitalization, transaction volume, and active developer communities. Together they span the principal architectural paradigms of distributed 
ledger technology along two largely orthogonal axes: Sybil resistance and 
leader selection and Byzantine agreement. This breadth provides a representative assessment of the broader ecosystem's PQ readiness.

\noindent\textbf{Our contributions.} We organise the study around four research
questions (RQ1--RQ4) and contribute:
\begin{itemize}
  \item a \emph{taxonomy} of the seven ledgers along three dimensions---%
  cryptographic primitives, consensus, and platform features---that frames the
  analysis (Section~\ref{sec:bc_apps});
  \item a \emph{vulnerability analysis} (RQ1) identifying, per system, which
  primitives are exposed and under which attack: online forgery,
  harvest-now-decrypt-later (HNDL), or consensus manipulation
  (\S~\ref{sec:vulnerabilities});
  \item a \emph{systematisation of defenses} (RQ2) separating, per platform, the
  safeguards deployed, the plans committed, and the directions still at the
  research stage (\S~\ref{sec:defense});
  \item a \emph{performance analysis} (RQ3) of substituting NIST PQ
  signatures---throughput, lost aggregation, storage footprint, and block-interval
  slack (\S~\ref{sec:challenges});
  \item an \emph{analysis of downstream effects} of PQ adoption (RQ4) on consensus convergence and
  miner incentives (\S~\ref{sec:shifts})
  \item a \emph{requirement set} for blockchain suitable PQ signatures (\S~\ref{sec:pq_requirements}) 
  
  
\end{itemize}

\noindent\textbf{Focus.} We focus only on L1 ledgers and their native
assets. We do not consider L2 rollups, bridges, permissioned chains, though we
note where bridges and rollups inherit the same exposure. We also do not consider wallet user experience and the overhead of PQ-TLS transport. Our quantitative
results are anchored to the NIST-standardised primitives and chain parameters of early 2026, before the final FIPS~206 standard and the Onramp second round.

\noindent\textbf{Summary of findings.} PQ adoption is not a drop-in replacement.
Larger signatures and keys inflate transaction size, which---under fixed block or
gas limits---cuts throughput, accelerates ledger growth, and slows propagation even
at an unchanged block interval; for aggregation-dependent designs such as Ethereum,
losing BLS aggregation is worse still, since per-slot signatures then scale with
the validator count. The binding constraint differs by chain, and the leading
mitigations---size-efficient lattice signatures, succinct SNARK/STARK aggregation,
and verification precompiles---each trade one cost for another.

\noindent\textbf{Outline.} Our paper is organized as follows. Section \ref{sec:background} provides background on quantum-resilient blockchain infrastructures. Section \ref{sec:bc_apps} surveys representative blockchain applications and their consensus and cryptographic foundations. Section \ref{sec:vulnerabilities} identifies vulnerabilities under the threat model defined in Appendix~\ref{sec:threat_model}. Section \ref{sec:defense} outlines emerging PQ defense strategies. Section \ref{sec:challenges} analyzes performance challenges in adopting PQ schemes, while Section \ref{sec:shifts} discusses expected downstream effects of PQ adoption. Section \ref{sec:pq_requirements} provides intuition on the properties of a blockchain suitable PQ signature. Section \ref{sec:open_prob} discusses open problems and future research directions with conclusion in Section \ref{sec:conclusion}.
\section{Background}
\label{sec:background}
In this section, we provide background necessary to contextualize the subsequent discussion on quantum-resilient blockchain infrastructures. We first outline NIST-standardized PQ cryptographic algorithms that support secure communication in a PQ era. We then examine SNARKs, a key primitive for scalable and privacy-preserving blockchain verification, whose constructions remain largely non-standardized and potentially quantum-vulnerable. Finally, we review consensus protocols, which define trust, security, and performance in blockchains.

\subsection{NIST Standardization of PQ Algorithms}


NIST's PQ standardisation effort, initiated in 2016, evaluates lattice-, code-, hash-, multivariate-, and isogeny-based candidates~\cite{nist-pqc-details}. The first FIPS standards were released in 2024~\cite{NIST-pqc-standard-Aug132024}: FIPS 203 (ML-KEM, derived from CRYSTALS-Kyber)~\cite{nist-fips203_encryption}, FIPS 204 (ML-DSA, derived from CRYSTALS-Dilithium)~\cite{nist-fips204_ds}, and FIPS 205 (SLH-DSA, derived from SPHINCS+)~\cite{nist-fips205_hds}. Falcon was selected for standardisation in 2022 and is being prepared as FIPS 206 (FN-DSA); as of this writing, NIST has not yet released an initial public draft. Public targets discussed by the IETF COSE working group anticipate IPD release in late 2026 or early 2027\cite{ietf-cose-falcon-04,nist-pqc-forum-fips206}. Its smaller signatures are particularly relevant to bandwidth-constrained settings such as blockchains. In March 2025, NIST selected Hamming Quasi-Cyclic (HQC) as a fifth, non-lattice KEM~\cite{nist2025hqc}.
 
NIST IR 8547~\cite{nistir8547} proposes deprecating 112 bit strength quantum vulnerable public-key algorithms (RSA-2048, ECC P-256) after 2030 and disallowing all quantum-vulnerable public-key algorithms---including EdDSA and $\geq$128-bit RSA/ECC variants, which skip the deprecation step---after 2035. The ``Onramp'' second-round signature standardisation explicitly seeks signatures with smaller signature/PK sizes and non-lattice security assumptions~\cite{nist-onramp-2024}---directly relevant to blockchains, where signature size dominates almost every performance metric.

\subsection{SNARK}
\label{sec:bk_snark}
SNARKs allow a prover to demonstrate the validity of a computational statement without revealing the witness or interacting with the verifier. They are characterised by \emph{succinctness} (short proofs, fast verification), \emph{non-interactivity} (typically via Common Reference String or Fiat--Shamir), and \emph{knowledge soundness} (only an entity that knows a valid witness produces an accepting proof). When the proof additionally hides the witness, the SNARK is \emph{zero-knowledge}. SNARKs support privacy-preserving payments (Zcash~\cite{zcash2016protocol}), verifiable rollups, and lightweight cross-chain verification.
 
In a quantum context, soundness and zero-knowledge are affected differently. Soundness in pairing-based SNARKs (Groth16~\cite{groth2016size}, KZG-based commitments~\cite{kzg2010polycommit}) reduces to elliptic-curve and pairing assumptions and is broken by Shor; the verifier of such a SNARK can be convinced of false statements once a CRQC is available. The zero-knowledge property of the same constructions is often statistical or perfect and survives a quantum adversary.
 
Two PQ directions for SNARKs have emerged. \emph{Hash-based} ones such as STARKs~\cite{ben2018scalable} achieve both soundness and zero-knowledge from collision-resistant hashing alone, at the cost of larger proofs. \emph{Lattice-based} SNARKs build soundness on Learning With Errors and Short Integer Solution assumptions~\cite{baum2018more, bootle2018sublinear}. Toolchains such as zkLLVM~\cite{nilfoundation_zkllvm} aim to make these backends interchangeable so that deployed circuits can migrate as PQ SNARK constructions mature.

\subsection{Consensus Protocols}
\label{sec:bk_consensus}

Consensus protocols are the foundation of blockchain systems, ensuring agreement on a shared ledger state among distributed nodes without central oversight. In adversarial and distributed environments, they guarantee safety, liveness, and fault tolerance through combinations of cryptographic proofs, economic incentives, and voting mechanisms. Their design governs the trade-offs among decentralization, security, and scalability---collectively, the blockchain trilemma. We review the principal consensus mechanisms used in contemporary blockchain networks, separating two largely orthogonal layers: leader and validator selection, which determines who proposes the next block, and Byzantine agreement, which determines how the network converges on it.
 
\subsubsection{Leader and Validator Selection}
\label{sec:bk_leader} 
The leader-selection layer determines which node is authorised to propose the next block and, in stake-based systems, which validators are eligible to participate in a given round. The principal mechanisms in use are as follows.

\noindent\textbf{Proof of Work (PoW).} Miners compute SHA-256 hashes until they find a nonce producing a digest below a target threshold. PoW underpins Bitcoin, Litecoin, Dogecoin, and Monero (using a memory-hard variant), and Ethereum~\cite{buterin2014ethereum} until its 2022 transition to PoS.
 
\noindent\textbf{Proof of Stake (PoS).} Validators lock native tokens; selection probability is proportional to stake; slashing penalises provably malicious behavior. \emph{Classical PoS} uses pseudorandom selection weighted by stake (Cardano/Ouroboros~\cite{kiayias2017ouroboros}; Polkadot's Nominated PoS~\cite{polkadot_vision_heterogeneous_multichain_2016}; Ethereum, which uses RANDAO with an in-research VDF backing~\cite{ethresearch_minimal_vdf}). \emph{Delegated PoS (DPoS)} delegates voting power to a small set of representatives that produce blocks on a schedule (EOS~\cite{eos2018}, TRON~\cite{tron2018whitepaper}, Steem~\cite{steem_whitepaper_2016}, Lisk~\cite{lisk_whitepaper_2018}). \emph{Pure PoS (PPoS)}, exemplified by Algorand, assigns proposer and committee roles via a Verifiable Random Function (VRF)~\cite{irtfCFRGvrlfDraft2019}: each round, validators evaluate the VRF locally on a public seed and self-select if their output falls below a stake-weighted threshold.
 
\noindent\textbf{Proof of Authority (PoA).} A fixed, identity-vetted validator set rotates on a deterministic schedule (VeChain~\cite{vechain_whitepaper_2018}, Energy Web Chain~\cite{energywebchain_paper_2020}; Ethereum testnets).
 
\noindent\textbf{Proof of History (PoH).} Solana's PoH is a verifiable, sequentially computed timestamp: each event is hashed together with the previous output via SHA-256, producing an unforgeable ordering. PoH is an iterated sequential function---related to but distinct from a VDF~\cite{boneh2018vdf}---whose security relies on the difficulty of computing SHA-256 chains faster than the leader, not on a hidden-order group. Solana pairs PoH with a stake-weighted leader schedule and the Tower BFT agreement protocol~\cite{yakovenko2018solana, shinobi_solana_primer}.
 
\subsubsection{Agreement and Finality}
\label{sec:bk_agreement}
 Once a block has been proposed, the agreement layer determines how the network reaches a binding decision on it and under what conditions that decision is considered final. The principal mechanisms are as follows.

\noindent\textbf{Longest-chain rule.} Used by Bitcoin and most PoW systems, this rule extends the chain with the greatest cumulative work. Finality is probabilistic: an adversary controlling 10\% of the hashrate has less than 0.03\% probability of reversing a six-block transaction~\cite{nakamoto2008bitcoin}.
 
\noindent\textbf{Practical Byzantine Fault Tolerance (PBFT).} PBFT and its descendants achieve deterministic finality among $3f+1$ replicas tolerating $f$ Byzantine faults via a three-phase commit~\cite{castro2002practical}. PBFT incurs $O(n^2)$ communication and is deployed primarily in permissioned systems (Hyperledger Fabric~\cite{hyperledger_fabric_distributed_os_2018}, Tendermint/Cosmos~\cite{kwon2014tendermint}, Quorum's Istanbul BFT~\cite{quorum2018whitepaper}). Algorand pairs PPoS with a one-shot Byzantine Agreement protocol that runs each round on a small, VRF-selected committee.
 
\noindent\textbf{Federated Byzantine Agreement (FBA).} FBA replaces fixed validator sets with overlapping quorum slices. Each node maintains a Unique Node List (UNL); consensus emerges when supermajorities of these lists intersect sufficiently. The Ripple Protocol Consensus Algorithm (RPCA)~\cite{schwartz2014ripple} and the Stellar Consensus Protocol~\cite{mazieres2015scp} are the canonical instances.
 
\noindent\textbf{Probabilistic BFT.} Snowflake, Snowball, Avalanche, Snowman \cite{teamrocket2019avalanche} --- the Snow* family uses repeated random subsampling with threshold-based vote tallies to achieve metastable convergence. Communication is near-constant per node; finality is probabilistic with exponentially decaying disagreement probability.
 
\noindent\textbf{Hybrid PoS--BFT finality gadgets.} Ethereum's post-Merge consensus combines LMD-GHOST (a fork-choice rule over the PoS-proposed chain) with Casper FFG (a finality gadget over checkpoint blocks)~\cite{buterin2019casperfriendlyfinalitygadget, buterin2020combiningghostcasper}. Block proposals occur every 12 seconds; finality is achieved at epoch boundaries ($\sim$6.4 minutes) via aggregated BLS attestations. The dependence on BLS aggregation is critical for performance and correspondingly critical for the PQ analysis.

\section{Blockchain Applications}
\label{sec:bc_apps}
In this section, we provide a brief overview of the seven prominent blockchain applications that we focus on, emphasizing their underlying consensus protocols and the cryptographic primitives that enable secure, decentralized operation. We summarize the results in Table \ref{tab:overview}.

\begin{table*}[t]
\caption{Cryptographic primitives and consensus design of surveyed blockchains. SC = smart contracts.}
\centering
\scriptsize
\setlength{\tabcolsep}{4pt}
\renewcommand{\arraystretch}{1.2}
\begin{tabular}{|l|>{\centering\arraybackslash}m{2.8cm}|>{\centering\arraybackslash}m{1.9cm}|>{\centering\arraybackslash}m{3.1cm}|>{\centering\arraybackslash}m{1.7cm}|>{\centering\arraybackslash}m{0.6cm}|>{\centering\arraybackslash}m{1.4cm}|}
    \hline
    & \multicolumn{2}{c|}{\textbf{Cryptographic Primitives}} & \multicolumn{2}{c|}{\textbf{Consensus}} & \multicolumn{2}{c|}{\textbf{Features}} \\
    \cline{2-7}
    \textbf{Chain} &
    \shortstack[c]{\textbf{Signature}\\\textbf{(usage)}} &
    \textbf{Hash} &
    \shortstack[c]{\textbf{Protocol}\\\textbf{(selection$\rightarrow$agreement)}} &
    \textbf{VDF/VRF} &
    \textbf{SC} &
    \shortstack[c]{\textbf{SNARK/}\\\textbf{ZK}} \\
    \hline
    Bitcoin   & ECDSA(secp256k1), Schnorr/BIP-340 (P2TR) & SHA-256, RIPEMD-160 & PoW $\rightarrow$ longest chain & -- & -- & -- \\
    \hline
    Ethereum  & ECDSA (tx), BLS12-381 (consensus), KZG & Keccak-256 & PoS $\rightarrow$ LMD-GHOST + Casper FFG & VRF/RANDAO & \checkmark & Yes \\
    \hline
    Algorand  & Ed25519 (tx), Falcon-1024 (state proofs) & SHA-256 & PPoS $\rightarrow$ Byzantine Agreement & VRF (ECVRF) & \checkmark & Yes \\
    \hline
    Solana    & Ed25519, W-OTS (opt-in vault) & SHA-256 (PoH), Keccak-256 & PoH leader sched.$\rightarrow$ Tower BFT & PoH (seq.\ hash) & \checkmark & -- \\
    \hline
    Avalanche & ECDSA(secp256k1) & SHA-256, RIPEMD-160, Keccak-256 (C-Chain) & PoS $\rightarrow$ Snowman (prob.\ BFT) & -- & \checkmark & -- \\
    \hline
    Monero    & Ed25519 (CLSAG ring sig.) & Keccak-256 & PoW (RandomX) $\rightarrow$ longest chain & -- & -- & Bulletproofs+ \\
    \hline
    XRPL      & Ed25519, ECDSA(secp256k1) (legacy) & SHA-512Half, RIPEMD-160 & UNL trust $\rightarrow$ RPCA (FBA) & -- & -- & -- \\
    \hline
\end{tabular}
\label{tab:overview}
\end{table*}

\subsection{Bitcoin}

Bitcoin~\cite{nakamoto2008bitcoin} is a PoW blockchain. Miners compete on SHA-256 hashing---using specialised hardware such as ASICs and GPUs---to find an output below the difficulty target and extend the longest chain. When a user wants to send funds, they create a transaction specifying the recipient address and the amount, and sign it with their private key. Signed transactions are broadcast, validated (signature, sufficient funds, no double-spend), pooled in the mempool, and included by miners in new blocks; six confirmations are considered final, giving an adversary at 10\% of network hashrate less than 0.03\% chance of reversal~\cite{nakamoto2008bitcoin}.

Bitcoin supports three main transaction output types, each with a different approach to storing the recipient's key material. \textbf{Pay-to-Public-Key (P2PK)} locks funds directly to the public key, which is stored on-chain in plaintext. \textbf{Pay-to-Public-Key-Hash (P2PKH)} stores only a hash of the public key; the key is revealed only when the owner spends the output. \textbf{Pay-to-Taproot (P2TR)}, activated in November 2021 via BIP-340/341/342~\cite{bip340, bip341, bip342}, introduces Schnorr signatures and encodes the public key directly in the locking script.

\noindent\textbf{Cryptographic primitives.} Bitcoin uses two signature schemes: ECDSA for legacy output types and Schnorr signatures (BIP-340) for P2TR outputs~\cite{bip341, bip342}. SHA-256 is used both for the PoW puzzle and for hashing public keys in P2PKH addresses, with RIPEMD-160 applied as an outer hash. Merkle trees allow any transaction in a block to be verified efficiently without downloading the whole block. Bloom filters let lightweight clients find their relevant transactions without revealing which addresses they own.

\subsection{Ethereum}

Ethereum~\cite{buterin2014ethereum} is a programmable blockchain supporting general-purpose smart contracts and the ecosystem of decentralised applications (dApps) built on them. It transitioned from PoW to PoS in September 2022 (the ``Merge''): validators stake ETH, and per 12-second slot one is pseudo-randomly selected to propose a block while the rest attest to the canonical chain; Casper FFG~\cite{buterin2019casperfriendlyfinalitygadget, buterin2020combiningghostcasper} finalises at epoch boundaries (${\sim}6.4$~min).

Two account types exist: \textit{Externally Owned Accounts} (EOAs), controlled by a private key, sign and send transactions; \textit{Contract Accounts} hold code that executes on invocation. Post-Pectra, EIP-7702~\cite{eip7702} lets EOAs delegate execution to contract code, blurring the distinction.




\noindent\textbf{Cryptographic primitives.} User transactions are authenticated with ECDSA over the secp256k1 curve. Validators use BLS signatures over the BLS12-381 curve for their attestations; a key property of BLS signatures is that thousands of individual attestations can be \textit{aggregated} into a single compact signature, which is critical for scalability: rather than storing tens of thousands of separate validator signatures per slot, the chain stores one aggregate~\cite{ethereum_pos_mechanism}. Keccak-256 is used for all general-purpose hashing, including transaction identifiers, block hashes, and address derivation. KZG polynomial commitments~\cite{kzg2010polycommit} are used in EIP-4844 to verify the availability of blob data for rollups. Randomness for validator selection is generated by RANDAO, a hash-chained beacon built from validators' contributions. Ethereum also supports zero-knowledge proofs through precompiled contracts for the BN254 pairing curve~\cite{eip196, eip197, eip1108}, enabling private transactions and verifiable rollups at L2.

\subsection{Algorand}

Algorand~\cite{gilad2017algorand} is a PoS blockchain with fast finality and a leaderless committee. A Verifiable Random Function (VRF) selects a different committee from all eligible accounts each round: each account evaluates the VRF locally on its private key and a public seed, and self-selects if the output falls below a stake-weighted threshold. No external coordination is needed, and no adversary knows the committee in advance---limiting the window for targeted attacks. A Byzantine Agreement protocol among the selected committee reaches finality in a single round with no forks~\cite{gilad2017algorand}. Smart contracts and stateless logic signatures run on the Algorand Virtual Machine (AVM).



\noindent\textbf{Cryptographic primitives.} Standard account transactions are signed with Ed25519. Committee selection uses an elliptic curve VRF (ECVRF)~\cite{irtfCFRGvrlfDraft2019} to produce a verifiable random output that determines participation.

Since 2022, Algorand has used Falcon-1024---a lattice-based signature scheme standardised by NIST---for \textit{state proofs}~\cite{algorand_postquantum_2022}: compact certificates, typically 600--900 KB, that attest to the chain's state over a 256-round window and are intended for cross-chain light-client verification. In September 2025, the go-algorand 4.3.0 upgrade introduced a \texttt{falcon\_verify} opcode into AVM v12~\cite{algorand_falcon_avm_2025}, allowing Falcon signatures to be verified directly on-chain. On November 3, 2025, the first quantum-resistant transaction was executed on Algorand's mainnet using a Falcon logic-signature template~\cite{algorand_falcon_tx_2025}. This deployment is opt-in: standard transactions still use Ed25519, and the consensus-layer VRF remains classical.

\subsection{Solana}

Solana~\cite{yakovenko2018solana} achieves sub-second block times by combining two mechanisms. PoH is a continuously running SHA-256 hash chain in which each output feeds the next; because computing it is inherently sequential, it serves as a cryptographic clock that lets validators agree on event ordering before any agreement runs. Tower BFT is a PoS-based agreement protocol layered on PoH that uses the historical record as a shared reference to cut messaging overhead~\cite{shinobi_solana_primer}. Leaders rotate on a stake-weighted schedule, and the Turbine protocol splits blocks into small packets propagated through a tree topology to handle Solana's high data throughput.



\noindent\textbf{Cryptographic primitives.} All standard transactions are signed with Ed25519. The PoH chain is built on iterated SHA-256 hashing. Solana offers an opt-in \textit{Winternitz Vault} as a quantum-resistant alternative for individual accounts~\cite{solana_pqc_vault_2025, solana_quantum_resistant_vault_2025}. The vault uses a Winternitz One-Time Signature (W-OTS) scheme: each spend generates a fresh one-time keypair, with the public key hashed via a truncated 224-bit Keccak-256 to derive a Program-Derived Address. Once an output is spent the vault is closed, preventing key reuse. This is a per-account opt-in feature and does not change the default Ed25519 signing scheme used by the rest of the network.

\subsection{Avalanche}

Avalanche~\cite{teamrocket2019avalanche} is a multi-chain platform: the \textit{Exchange Chain} (X-Chain) handles asset creation and trading, the \textit{Platform Chain} (P-Chain) manages validator staking, subnet creation, and cross-chain coordination, and the \textit{Contract Chain} (C-Chain) is an Ethereum-compatible execution environment. Consensus runs the \textit{Snowman} protocol (Snow* family): each validator repeatedly polls a small random sample of peers about their preferred block and updates its preference when a sufficient fraction agrees. Repeated sampling with threshold voting converges on a single block without all-to-all communication; finality is probabilistic, with disagreement probability decaying exponentially per round. Avalanche also supports \textit{subnets}: custom blockchains with their own validator sets, VMs, and governance, tailored to specific applications.




\noindent\textbf{Cryptographic primitives.} All three built-in chains use ECDSA over secp256k1 for transaction signing. X-Chain and P-Chain derive addresses by applying SHA-256 followed by RIPEMD-160 to the public key, producing a 20-byte address. The C-Chain follows Ethereum's convention, deriving addresses from a Keccak-256 hash.

\subsection{Monero}

Monero~\cite{noether2016ringct-mrl0005} is a privacy blockchain: transaction amounts, senders, and recipients are all hidden by default, so no outside observer can link a transaction to a user or determine an amount. Consensus is PoW under RandomX, a memory-hard algorithm optimised for general-purpose CPUs to resist ASIC dominance; P2Pool supports decentralised cooperative mining.

Privacy rests on three complementary mechanisms. \textbf{Ring signatures} let a sender sign on behalf of a group of possible signers (a \textit{ring}); Monero currently uses CLSAG~\cite{goodell2020clsag} (which superseded MLSAG), and a per-spend \textit{key image} prevents double-spending without revealing the signer. \textbf{Stealth addresses} derive a fresh one-time destination from the recipient's public key, so transactions to the same recipient are unlinkable on-chain. \textbf{RingCT}~\cite{noether2016ringct-mrl0005} hides amounts via Pedersen commitments---which bind a value without revealing it---and Bulletproofs+~\cite{bunz2018bulletproofs} range-prove non-negativity of committed amounts.

\noindent\textbf{Cryptographic primitives.} Spend and view keys are derived on the Ed25519 curve and combined via Keccak-256 to produce one-time stealth addresses. CLSAG ring signatures, Pedersen commitments, and Bulletproofs+ all rely on ECC.

\subsection{XRP Ledger}

The XRP Ledger (XRPL)~\cite{schwartz2014ripple} is a blockchain for fast, low-cost payments and asset exchange, using neither PoW nor PoS. Consensus is the \textit{Ripple Protocol Consensus Algorithm} (RPCA), an instance of Federated Byzantine Agreement~\cite{mazieres2015scp}: each node keeps a \textit{Unique Node List} (UNL) of trusted validators, and consensus fires when a supermajority of that UNL agrees on the same transaction set. Sufficient overlap across nodes' UNLs converges the network on a single ledger state each round. XRPL records both successful and failed transactions; per-account sequence numbers enforce strict ordering (a transaction is accepted only if its sequence matches the account's next expected value), preventing replay and reordering, and a small XRP reserve deters spam.



\noindent\textbf{Cryptographic primitives.} XRPL supports two signature schemes: Ed25519 (the default since 2017) and ECDSA over secp256k1 (retained for legacy compatibility)~\cite{xrpl_cryptographic_keys}. Transaction and ledger data is hashed with SHA-512Half---the first 256 bits of a SHA-512 output---which provides efficiency on 64-bit hardware. Account addresses are encoded in Base58 from a RIPEMD-160 hash of the public key.

\section{Threat Model}
\label{sec:threat_model}
 
The adversary $\mathcal{A}$ has access to a \textbf{cryptographically relevant quantum computer (CRQC)}: a fault-tolerant quantum machine large enough to run Shor against 256-bit elliptic curves and Grover against 256-bit hash inputs. $\mathcal{A}$ additionally controls polynomial classical computation, observes all on-chain data and peer-to-peer traffic, and can selectively delay or reorder messages within the underlying network model. $\mathcal{A}$ cannot break information-theoretic primitives or PQ schemes whose underlying lattice, code, isogeny, or hash assumptions remain true.
 
Our threat model defines a CRQC by \emph{capability} rather than by hardware parameters; quantitative attack-window claims throughout the paper inherit assumptions from the resource estimates we cite. Webber et al.~\cite{webber2022impact} finds that breaking secp256k1 requires $\sim$${317 \times 10^6}$ physical qubits for a one-hour attack and $\sim$${13 \times 10^6}$ for a 24-hour attack, assuming a surface code with 1~$\mu$s code cycles, 10~$\mu$s reaction time, and $10^{-3}$ gate error rate. Litinski~\cite{litinski2023} cuts the per-key Toffoli count to $\sim$${5 \times 10^7}$ and, via an active-volume architecture with non-local inter-module connections, reduces the spacetime cost by a factor of 300--700 relative to a 2D-local baseline---bringing the qubit requirement to $\sim$${6.9 \times 10^6}$ (6{,}000 modules of 1{,}152 qubits each) for one 256-bit ECC key every 10 minutes, at roughly half Webber et al.'s 24-hour qubit count. Ongoing work with qLDPC codes and improved magic-state factories continues to compress these estimates~\cite{gidney2025factor}.

These figures are not properties of a CRQC per se but of specific (qubit count, gate fidelity, cycle time, connectivity) tuples, and different methodological choices lead to markedly different conclusions. Aggarwal et al.~\cite{aggarwal2017quantumbitcoin}, using an optimistic-projection approach, argue the 10-minute attack window could be reached as early as 2027. Readers should treat the attack-window claims here as anchored to the parameter sets cited above, not as consensus estimates.
 
The available attack window determines which primitives are at risk. We consider the attacks below.

    \textbf{Online forgery.} $\mathcal{A}$ observes a public key revealed by a transaction and must derive the private key before the transaction is finalised. The window is the block confirmation interval. Mitigations include address-reuse hygiene and short confirmation times.
    
    \textbf{Harvest-now-decrypt-later (HNDL) and long-term integrity.} $\mathcal{A}$ records on-chain data today and attacks it once a CRQC is available. This threatens P2PK addresses, exposed validator BLS keys, and any privacy primitive whose computational hiding rests on classical assumptions (e.g., Bulletproof range proofs, ECC-based ring signatures protecting historical Monero transactions). Mitigation requires either that the data never be exposed in raw form or that legacy data be migrated before $\mathcal{A}$ becomes capable.
    
    \textbf{Consensus manipulation.} $\mathcal{A}$ forges VRF outputs to bias validator selection, forges VDF or state-proof signatures to attest to a forged ledger state, or recovers BLS aggregate-signing keys to fabricate quorums. A single successful forgery here can compromise an entire epoch.

The relevant window for online forgery is chain-specific: ~10 minutes for Bitcoin, 0.4 s for Solana etc. For sub-second-finality chains, online forgery is not currently considered feasible under any published resource-estimate baseline; the dominant quantum threat for these chains is HNDL of validator/account keys whose public keys are reused across many transactions.
\section{Vulnerability to Quantum Adversaries}
\label{sec:vulnerabilities}

Based on the threat model in Appendix~\ref{sec:threat_model}, we now 
examine each blockchain individually. We address \textit{RQ1: What 
cryptographic building blocks of current blockchain designs make them a 
target in the presence of a quantum adversary?} For each system we identify 
the vulnerable components and the attack scenario---online forgery, 
harvest-now-decrypt-later, or consensus manipulation---under which each is 
exposed. A summary appears in Table~\ref{tab:solutions}.

\subsection{Bitcoin}

Bitcoin's quantum exposure depends on which output type holds the funds.

\noindent\textbf{P2PK} exposes the recipient's public key permanently, so any future CRQC can derive the private key and drain the funds---no spend required. Many Satoshi-era coinbase outputs remain in this format.
 
\noindent\textbf{P2PKH} reveals the key only when the owner spends, providing meaningful long-term protection but leaving a residual online-forgery window between broadcast and confirmation. Under optimistic hardware assumptions, Aggarwal et al.~\cite{aggarwal2017quantumbitcoin} estimate a key derivation under 10 minutes by 2027---comparable to the average block interval; conservative assumptions extend the window but do not close it.
 
\noindent\textbf{P2TR} (November 2021) encodes a 32-byte tweaked Schnorr key directly in the locking script. \emph{Key-path spends}---the common case for single-signature wallets---expose the key from first deposit, matching P2PK's long-term vulnerability. \emph{Script-path spends} hash-commit the script tree and defer key exposure until a non-default branch executes, and are correspondingly less exposed. Since most P2TR usage today is key-path, the long-term characterisation applies to the bulk of P2TR value.
 
Growing P2TR adoption (Ordinals, BRC-20 tokens, modern wallets) continuously expands the pool of exposed keys. Migrating all UTXOs to quantum-resistant addresses is bound below at 76.16 days of full-bandwidth network processing~\cite{pont2024bitcoindowntime}, roughly doubling to ${\sim}152$ days when 50\% of block capacity is reserved for regular traffic---both operationally untenable for an L1 of Bitcoin's size.
 
PoW itself is comparatively resilient: Grover's algorithm offers only a quadratic speedup on hash search, and ASIC clock speeds outpace projected near-term quantum hardware by orders of magnitude~\cite{aggarwal2017quantumbitcoin}.

\subsection{Ethereum}
 
Ethereum's quantum exposure spans three distinct layers.
 
\noindent\textbf{Transaction layer.} Accounts sign with ECDSA; because addresses are Keccak-256 hashes of the public key, the key is exposed only from broadcast through confirmation---the same online-forgery window as Bitcoin P2PKH. Post-Pectra, EIP-7702-delegated EOAs retain ECDSA authorisation; delegating to contract code does not confer PQ protection on the underlying signing key.
 
\noindent\textbf{Consensus layer.} Validator BLS attestation keys are reused across rounds and publicly known, exposing them to HNDL today. A CRQC that recovers one can forge attestations for that validator's stake.
 
\noindent\textbf{Data-availability layer.} KZG polynomial commitments (EIP-4844) are pairing-based, so a CRQC can produce commitments that appear valid but correspond to data never published, undermining rollups that inherit DA from EIP-4844.
 
Ethereum's L2 rollup ecosystem (Groth16, PLONK) uses pairing-based ZK proofs whose soundness Shor also breaks; the ZK property is typically unconditional and survives.

\subsection{Algorand}
 
Algorand's quantum exposure divides cleanly between two layers.
 
\noindent\textbf{Standard transactions} are signed with Ed25519. The public key is included in each transaction, so it is exposed on first use---similar to Ethereum, the key is visible and the account is vulnerable to online forgery from that point.
 
\noindent\textbf{Committee selection} uses a VRF to privately determine which accounts are eligible to participate in each consensus round. The VRF relies on an EC private key. An adversary that recovers a validator's VRF key can predict or manipulate which accounts appear to be selected, skewing the committee composition and potentially influencing consensus outcomes.
 
Algorand's state proof subsystem already uses Falcon-1024, a PQ signature scheme, for cross-chain attestations. However, this does not protect standard user transactions or the VRF-based committee selection, which remain the dominant quantum exposure.
 
\subsection{Solana}
 
Solana's primary quantum exposure is its use of Ed25519 for all standard transaction signatures, including the keys that validators use to participate in the network. These keys are publicly known and are reused across many transactions, making them targets for HNDL attacks.
 
PoH, Solana's verifiable timestamp mechanism, is built on iterated SHA-256 hashing. As discussed above, SHA-256 retains adequate security against Grover's algorithm. PoH's sequentiality does not rely on a group structure that Shor can exploit, so PoH itself is not directly threatened by quantum computing. The attack surface is the Ed25519 signature layer, not the timekeeping mechanism.
 
The optional Winternitz Vault on Solana, offers per-account quantum resistance, but the vast majority of accounts and all consensus-layer keys remain on Ed25519.
 
\subsection{Avalanche}
 
All three of Avalanche's built-in chains (X-Chain, P-Chain, C-Chain) use ECDSA over secp256k1, broken by Shor's algorithm. On the X-Chain and P-Chain, addresses are derived from a hash of the public key, mirroring P2PKH's partial protection. The C-Chain follows Ethereum's address derivation and carries the same transaction-layer exposure.
 
Validator keys on Avalanche's P-Chain are long-lived and reused across many rounds; their public keys are known to the network. A quantum adversary that recovers a validator's private key can forge that validator's attestations, claiming stake-proportional influence over consensus. Avalanche's Snowman probabilistic BFT converges quickly (in about 1--2 seconds), but this short confirmation window is a probabilistic mitigation, not a structural defense against a quantum key-recovery attack.
 
\subsection{Monero}
 
Monero's quantum exposure is more nuanced than in account-based chains, because its privacy model distributes the risk differently across its cryptographic components.
 
\noindent\textbf{Ring signatures} (CLSAG~\cite{goodell2020clsag}) hide the actual signer among a group of possible signers. Signer ambiguity is \emph{computational}, not information-theoretic---it rests on DDH over Curve25519, which Shor breaks. A CRQC therefore de-anonymises CLSAG for any ring whose members' public keys it has observed by recovering each member's secret scalar and checking which key was used. Unforgeability likewise reduces to ECC and falls to Shor. The theft path is: (i)~Shor recovers the victim's private spend key from any observed public key on the same curve (e.g., a stealth-address derivation, below); (ii)~the recovered key lets the adversary compute the correct \emph{key image} for a controlled output; (iii)~the adversary signs a forged CLSAG with that key image, passing Monero's double-spend check.
 
\noindent\textbf{Stealth addresses} derive one-time destinations from the recipient's public key via ECC. A CRQC recovers the recipient's spend key from any observed stealth address, linking transactions and enabling theft.
 
\noindent\textbf{Pedersen commitments} hide amounts unconditionally, so historical amount confidentiality survives quantum. Their \textit{binding} property, however, reduces to ECDLP: a CRQC can open the same commitment to two values, forging balance-preserving but invalid transactions that still verify.
 
\noindent\textbf{Bulletproofs+}~\cite{bunz2018bulletproofs} range-prove non-negativity. The ZK property is statistical and survives; soundness rests on ECC and falls to Shor, letting a CRQC produce range proofs for negative or otherwise invalid amounts.

\subsection{XRP Ledger}
 
XRPL transactions are signed with Ed25519 (default) or ECDSA (legacy), both broken by Shor. Accounts are addressed as a RIPEMD-160 hash of the master public key, so funds at never-signed addresses retain hash-based protection. However, every signed transaction reveals the master public key on-chain, and the same key is reused for every subsequent transaction from that account. Once revealed, all past and future transactions from that account are quantum-exposed until the key is rotated or the account is deleted.
 
Validator messages in RPCA are also signed with Ed25519. A CRQC that recovers a validator's key can forge UNL attestations, fabricating agreement among trusted validators and forcing the network to accept an invalid ledger state.
 
SHA-512Half, used for transaction and ledger hashing, retains adequate security under Grover.

\section{Defenses Against Quantum Adversaries}
\label{sec:defense}

In response to vulnerabilities to quantum adversaries, a growing body of research and development has focused on designing mechanisms that remain resilient in a PQ setting. In this section we answer the question \textit{RQ2: What solutions have been proposed by blockchain applications to tackle a quantum adversary?} For each system we distinguish \emph{deployed safeguards} (in production, including opt-in features), \emph{committed transition plans} (with a public specification or implementation milestone), and \emph{research directions} (academic or working-group proposals not yet integrated). A summary appears in Table~\ref{tab:solutions}.

\begin{table*}[t]
\caption{PQ-Transition Posture of Surveyed Blockchains.}
\centering
\scriptsize
\renewcommand{\arraystretch}{1.3}
\resizebox{\textwidth}{!}{%
\begin{tabular}{|l|>{\centering\arraybackslash}m{3.6cm}|>{\centering\arraybackslash}m{2.6cm}|>{\centering\arraybackslash}m{4cm}|>{\centering\arraybackslash}m{4.6cm}|}
  \hline
  \textbf{Blockchain} &
  \textbf{Quantum-Vulnerable Primitives} &
  \textbf{Deployed PQ Safeguards} &
  \textbf{Committed Transition Plans} &
  \textbf{Research Directions} \\
  \hline

  Bitcoin &
  ECDSA, Schnorr (P2TR) &
  None &
  None at protocol level &
  Bitcoin Post-Quantum (BPQ) hard-fork prototype using XMSS; BIP-360 (P2QRH) draft for quantum-resistant outputs \\
  \hline

  Ethereum &
  ECDSA (tx), BLS12-381 (consensus), KZG, ECC-VRF/VDF beacon &
  None &
  Roadmap toward W-OTS / zk-STARK-based execution layer &
  Account-abstraction Falcon verifier; Falcon precompile (EIP-7592); SNARK aggregation of PQ signatures \\
  \hline

  Algorand &
  Ed25519 (default tx), ECVRF (committee selection) &
  Falcon-1024 in state proofs (since 2022); opt-in Falcon-1024 logic-sigs via \texttt{falcon\_verify} AVM opcode (Sept.\ 2025; first PQ tx Nov.\ 2025) &
  Native Falcon transaction support proposed; PQ VRF integration under research &
  X-VRF (broken in 2024), LB-VRF (lattice-based, scaling-limited) \\
  \hline

  Solana &
  Ed25519 (default tx, validator stake keys) &
  Opt-in W-OTS ``Winternitz Vault'' (non-default) &
  None at protocol level &
  None publicly committed beyond the Vault \\
  \hline

  Avalanche &
  ECDSA(secp256k1) across X/P/C-Chains &
  None &
  None at protocol level &
  Lattice-based signatures explored by Ava Labs; subnet-level PQ pilots possible \\
  \hline

  Monero &
  Ed25519 (CLSAG, stealth addresses), DLP-based commitments and Bulletproofs+ &
  None &
  FCMP++ proposal under evaluation for a future hard fork &
  Lattice RingCT, MatRiCT, ChipmunkRing, GLYPH; PQ ring signatures more broadly \\
  \hline

  XRPL &
  Ed25519, ECDSA(secp256k1) &
  None &
  XRPL Standards \#79 / \#295 community discussions on lattice-based signature integration &
  Ripple--Trinity College Dublin (ADAPT) collaboration on PQ signatures and ZKPs \\
  \hline
\end{tabular}%
}

\label{tab:solutions}
\end{table*}

\subsection{Bitcoin}

Bitcoin has no PQ defenses deployed and none committed at the protocol level. The only operative safeguard is user discipline: spending from P2PKH or P2WPKH outputs with non-reused addresses limits HNDL exposure, at the cost of a residual online-forgery risk during transaction confirmation, whereas P2TR key-path usage expands HNDL exposure rather than reducing it (\S~\ref{sec:vulnerabilities}).
 
\noindent\textbf{Research Directions.} Two proposals target a quantum-resistant Bitcoin. BIP-360 (P2QRH)~\cite{bip360} drafts a new SegWit output type whose locking script commits to a quantum-resistant public-key hash and accepts PQ signatures (initially Falcon and SLH-DSA) via soft fork, providing forward-compatible PQ-protected sends without migrating legacy UTXOs. The Bitcoin Post-Quantum (BPQ) prototype~\cite{bitcoinpq2025}, a hard-forked variant live since December 2018, instead implements XMSS over W-OTS+; because XMSS signatures are several times larger than ECDSA, BPQ raises the block-weight limit from 4M to 32M units and replaces SHA-256 PoW with Equihash~\cite{biryukov2016equihash}, a generalised-birthday puzzle that admits less quantum speedup than preimage search, though it remains a niche proof-of-feasibility with negligible hashrate.

\subsection{Ethereum}

The Merge (2022) eliminated Ethereum's PoW exposure to Grover~\cite{ethereum_futureproofing} but left its ECDSA, BLS, KZG, and beacon-VRF/VDF dependencies intact. Ethereum's committed direction is the published roadmap toward a W-OTS- and zk-STARK-based execution layer, which targets PQ security with transparency and no trusted setup~\cite{btq_eth_roadmap}; as an early demonstration, BTQ and StarkWare verified a Falcon signature on StarkNet, an Ethereum Layer-2 network~\cite{btq_falcon_starkware}.
 
\noindent\textbf{Research Directions.} The concrete transaction-level mechanisms remain under investigation. An account-abstraction (EIP-4337) Falcon verifier can be deployed as a contract~\cite{ethresearch_aa_falcon, ethresearch_falcon_gnarly} but is gas-prohibitive in EVM bytecode, motivating the proposed EIP-7592 Falcon precompile~\cite{eip_7592_falcon}; an execution-layer Ethereum Object Format predicate would instead admit arbitrary validation code at the cost of new denial-of-service surfaces~\cite{quantuminsider_2024, ethereum_futureproofing}. The hardest requirement is replacing BLS attestation aggregation, for which no native PQ analogue offers comparable size guarantees (\S~\ref{sec:agg_loss}): the Ethereum Foundation's roadmap targets STARK-friendly hash-based multi-signatures~\cite{drake2025hashbased, btq_eth_roadmap}, Privacy~\&~Scaling Explorations has aggregated Falcon signatures via LaBRADOR-style lattice arguments~\cite{pse_falcon_aggregation}, and hybrid-cost models bound the per-transaction overhead of partial measures~\cite{campbell2025hybrid}.

\subsection{Algorand}

Algorand maintains the most advanced PQ posture of the surveyed networks, and its safeguards already span two layers. Falcon-1024 has secured its state-proof attestations since 2022~\cite{algorand_postquantum_2022, algorand_technology_postquantum}, and the deployment has since reached the transaction layer: the go-algorand 4.3.0 consensus upgrade activated AVM v12 with the \texttt{falcon\_verify} opcode in September 2025~\cite{algorand_falcon_avm_2025}, and on November 3, 2025 the Foundation executed the first quantum-resistant mainnet transaction via a Falcon logic-signature template~\cite{algorand_falcon_tx_2025}. Native, default Falcon transaction support is the committed next step but has not yet landed in the reference protocol; as \S~\ref{sec:challenges} shows, native Falcon-1024 roughly halves the throughput envelope of Falcon-512 under the same block parameters, which constrains the design space for any future native-PQ rollout.
 
\noindent\textbf{Research Directions.} The principal unmet dependency is a scalable PQ replacement for the consensus-layer VRF. The X-VRF construction was shown insecure under realistic adversarial assumptions in 2024~\cite{bodaghi2024_breakingXVRF}, and the lattice-based LB-VRF---an 84-byte output with a $\sim$5~KB proof and millisecond evaluation and verification---has not scaled beyond $\sim$1{,}000 nodes owing to its $\sim$8~MB-per-instance overhead and few-time-key constraint~\cite{esgin2021_practicalPostQuantumFewTimeVRF}. A scalable PQ VRF therefore gates PQ-complete Algorand consensus.

\subsection{Solana}
 
Solana's only deployed safeguard is the optional Winternitz Vault~\cite{solana_pqc_vault_2025, solana_quantum_resistant_vault_2025}, a per-account W-OTS defense against HNDL rather than a network-wide transition; it is also constrained by Solana's 1{,}232-byte serialized transaction limit (\S~\ref{sec:challenges}), which restricts the kinds of PQ schemes expressible as native transaction signatures. On the committed side, SIMD-0296 proposes raising that cap from 1{,}232~B to 4{,}096~B via the v1 transaction format (SIMD-0385)~\cite{solana_simd0296}, widening the set of PQ signatures that could be carried natively.
 
\noindent\textbf{Research Directions.} Beyond the opt-in Vault and the proposed cap increase, no PQ scheme is under active protocol-level development; public engineering statements indicate that scalability has taken priority over PQ migration~\cite{solana_quantum_strategy_2025}, leaving native PQ transaction signatures as future work.
 
\subsection{Avalanche}
Ava Labs has acknowledged the quantum risk to its ECC-only stack but has deployed and committed to nothing at the protocol level.
 
\noindent\textbf{Research Directions.} Ava Labs' public position is that lattice-based signatures are the leading candidate replacement~\cite{ava_labs_lattice_2024}; because such signatures are roughly an order of magnitude larger than ECDSA---with corresponding bandwidth and storage costs---the team has not committed to a specific scheme or fork timeline, and its configurable subnets are a plausible venue for early PQ pilots that need not commit the whole tri-chain architecture.
 
\subsection{Monero}
 
Monero has no PQ safeguard deployed, but its six-month hard-fork cadence positions it to absorb one relatively quickly~\cite{monero_quantumcanary_2025}, and a 2020 community review established the groundwork~\cite{monero_pqc_2020proposal}. Its committed architectural direction is FCMP++~\cite{parker2024_fcmpplusplus}, the proposal under most active development: it supports full-chain membership proofs, spend authorisation, and linkability while preserving full-set anonymity, and adds transaction chaining, outgoing view keys, and forward secrecy. FCMP++ is the scaffold into which PQ primitives must eventually be slotted, but it is not yet PQ-secure end-to-end. The earlier Seraphis proposal~\cite{koe2022seraphis} improves transaction modularity by separating membership proofs from ownership and balance proofs, yet offers only limited PQ guarantees because its primitives remain ECC-based~\cite{monero_seraphis_2021, monero_quantumcanary_2025}.
 
\noindent\textbf{Research Directions.} A PQ-secure Monero requires a primitive class that preserves signer ambiguity, which single-signer NIST schemes cannot provide (\S~\ref{sec:vulnerabilities}). Candidate PQ ring constructions that could plug into a future RingCT redesign---Lattice RingCT over Ring-SIS~\cite{torres2018lrct}, MatRiCT~\cite{matrict} (with $\sim$23~ms generation and verification and proofs two orders of magnitude smaller than earlier PQ designs), the hash-based GLYPH~\cite{cryptoeprint:2017/766}, and the lattice-based ChipmunkRing~\cite{chipmunkring2025} ($\sim$20--280~KB signatures for rings of 2--64 members)---exist, but none has been integrated into a Monero release, and no single construction yet unifies linkable PQ ring signatures, PQ-binding commitments, and PQ range proofs.
 
\subsection{XRP Ledger}

XRPL has no protocol-level PQ deployment in the reference \texttt{rippled} implementation, but a community standards process is under way: the Foundation has opened formal Standards discussions \#79 (PQ-readiness) and \#295 (PQ signature support)~\cite{xrplf2024pqreadiness, xrplf2025pqsignatures}, framing migration options for lattice-based transaction signatures and validator keys, and independent analyses have mapped concrete migration paths~\cite{taghi2025pq}.
 
\noindent\textbf{Research Directions.} The supporting cryptographic research is being developed through a collaboration with Trinity College Dublin's ADAPT Centre on PQ signatures, ZKPs, and validator infrastructure, funded under Ripple's University Blockchain Research Initiative~\cite{ripple2024quantum, malhotra2025xrpl}, advancing the building blocks on which the standards discussions above ultimately depend.

\section{Performance Analysis of PQ Adoption}
\label{sec:challenges}
 
As blockchain systems transition to PQ signature schemes, a key concern is the trade-off between enhanced cryptographic security and system efficiency. This section addresses \textit{RQ3: What is the impact on performance when existing blockchain applications shift to a PQ signature scheme?} We cover four axes: transaction throughput, loss of signature aggregation, public-key and storage footprint, and block-interval timing.
 
We benchmark three NIST-standardised PQ signatures: \textbf{ML-DSA-44} (FIPS~204, lattice-based), \textbf{Falcon-512} (FIPS~206 draft, lattice-based), and \textbf{SLH-DSA-128f} (FIPS~205, hash-based).

\subsection{Impact on Transaction Throughput}
\label{sec:throughput}
 
The \textit{theoretical TPS} of a blockchain is the maximum number of minimum-sized transactions that fit in one block, divided by the block interval:
\[
\text{Theoretical TPS} = \frac{\text{Maximum Block Size}}{\text{Minimum Tx Size} \times \text{Block Time}}.
\]
Replacing a classical signature with a larger PQ signature increases the minimum transaction size and reduces TPS in proportion. Per-chain results are summarised in Table~\ref{tab:blockchain_tps_pq}.

\noindent\textbf{Bitcoin.}
A 1-input/1-output P2WPKH transaction is $\approx$109.5~virtual bytes~(vB), with the witness (71-byte ECDSA sig + 33-byte PK = 104~B~$\approx$~26~vB) benefiting from the SegWit discount. With Bitcoin's 1{,}000{,}000-vB block cap and 600-second block interval, classical theoretical TPS is $\approx$15. Under ML-DSA-44, the witness grows to 3{,}732~B ($\approx$933~vB), bringing the transaction to $\approx$1{,}016.5~vB and TPS down to $\approx$1.6---an \textbf{89\% reduction}. Falcon-512 is less severe, with a 1{,}563-B witness ($\approx$391~vB) yielding a $\approx$474-vB transaction and TPS of $\approx$3.5 ($\sim$77\% reduction). SLH-DSA-128f is the most damaging: its 17{,}088-byte signature pushes the transaction to $\approx$4{,}364~vB and TPS to $\approx$0.4 ($\sim$97\% reduction); its 32-byte PK is essentially unchanged from ECDSA's 33~bytes, so the penalty comes almost entirely from the signature.

\noindent\textbf{Ethereum and Avalanche C-Chain.}
Both chains use Ethereum's gas model. Transactions pay a 21{,}000-gas intrinsic base; calldata is priced at 16~gas per nonzero byte under EIP-2028~\cite{eip_2028}, rising to a 40~gas/nonzero-byte floor for calldata-heavy transactions under EIP-7623 ~\cite{eip_7623_calldata}. Ethereum's per-block gas limit was raised to 60M in November~2025\cite{ethereum_60M}, giving a classical TPS of $\approx$238 at a 12 s slot.

Avalanche's C-Chain is governed by ACP-176/Octane~\cite{avalanche_acp176_octane},
which replaces the previous fixed per-block gas limit with a dynamic
\emph{target gas-consumption rate}~$T$, validator-signalled and adjusted
over time. The initial target at activation was $T = 1.5$M~gas/sec
(15M~gas per 10-second rolling window); subsequent validator signalling
has raised this by ${\sim}30\%$ to $T \approx 2.1$M~gas/sec as of
mid-2025~\cite{avalanche_octane_gastarget}. A per-block soft cap of
$\min(8T, 15\text{M})$~gas limits burst-block consumption, but
\emph{sustained} throughput is determined by $T$. At
$T = 2.1$M~gas/sec, the C-Chain's classical theoretical TPS is therefore
${\approx}100$, in line with its observed real-time throughput
(Appendix~A). A burst cap of 15M gas allows the C-chain to yield $\approx$476~TPS at a $\approx$1.5-s slot.
 
PQ adoption affects the gas cost in two ways. First, the signature appears as calldata and triggers the EIP-7623 floor: an ML-DSA-44 signature costs $21{,}000 + 2{,}420 \times 40 = 117{,}800$~gas intrinsically. Second, in-EVM PQ verification is prohibitively expensive, so practical deployment requires a precompile (e.g., EIP-7592~\cite{eip_7592_falcon}); its gas cost is not yet finalised, so Table~\ref{tab:blockchain_tps_pq} reports calldata-bounded upper limits assuming negligible precompile cost. The bounds further assume the PQ public key is stored once in an account-abstraction wallet (per EIP-4337~\cite{ethresearch_aa_falcon}) so that per-transaction calldata carries the signature only; including the PK on every transaction tightens the bounds by roughly 1.5--2$\times$ for lattice schemes, while SLH-DSA---with its 32-byte PK---is unaffected. Under the AA-wallet assumption, Ethereum sustains at most $\approx$42, $\approx$105, and $\approx$7~PQ-tx/s for ML-DSA-44, Falcon-512, and SLH-DSA-128f respectively; Avalanche yields $\approx$85, $\approx$210, and $\approx$14.

\noindent\textbf{Solana.}
Solana enforces a 1{,}232-byte cap on serialised transaction size. ML-DSA-44 signatures alone (2{,}420~B) exceed this cap, as do all SLH-DSA-128f parameter sets. Falcon-512's 666-byte signature fits, but adding the 897-byte public key pushes the combined payload over the limit. None of the three NIST PQ schemes can serve as a drop-in replacement without changes to Solana's transaction format. According to Solana's roadmap SIMD-0296 increasing serialised transaction cap to 4{,}096~B
would admit Falcon-512 (sig+PK ${\approx}1.6$~KB) but still not
ML-DSA-44 or SLH-DSA-128f.

\noindent\textbf{Monero.}
Monero's privacy depends on ring signatures, which obscure the actual signer among a group of possible signers. ML-DSA-44, Falcon-512, and SLH-DSA-128f are ordinary single-signer schemes; substituting any of them would destroy anonymity guarantees. A PQ-compatible Monero requires a different primitive class entirely---such as a lattice-based linkable ring signature (e.g., Lattice~RingCT~\cite{torres2018lrct}, MatRiCT~\cite{matrict}, or ChipmunkRing~\cite{chipmunkring2025})---whose size and verification costs remain research-stage. Because of this, the throughput and storage figures we report for Monero in Tables~\ref{tab:blockchain_tps_pq} and~\ref{tab:blockchain_storage_pq} are not direct PQ-substitution projections but illustrative bounds assuming a single-signer NIST scheme were forced into the slot of CLSAG (a configuration that would not be deployed in practice).

\noindent\textbf{Algorand.}
Algorand's 5~MiB blocks\cite{algorand_consensus} close every 2.82 s. The average transaction is $\approx$200~bytes, with a 64-byte Ed25519 signature and 32-byte PK accounting for 96~bytes; classical TPS is $\approx$9{,}295. ML-DSA-44 inflates the sig+PK to 3{,}732~B, growing the transaction to $\approx$3{,}836~B and dropping TPS to $\approx$484 ($\sim$95\% reduction). Falcon-512 is more manageable at 1{,}563~B sig+PK, yielding a $\approx$1{,}667-B transaction and TPS of $\approx$1{,}115 ($\sim$88\% reduction); note, however, that Algorand's deployed scheme is Falcon-1024 (sig 1{,}280~B, PK 1{,}793~B), which would yield a $\approx$3{,}177-B transaction and a correspondingly lower TPS of $\approx$585. SLH-DSA-128f, at 17{,}120~B sig+PK, brings TPS to $\approx$108---a $\sim$99\% reduction. Algorand's existing deployment of Falcon-1024 in state proofs makes Falcon the most realistic path for native PQ transactions.

\noindent\textbf{XRPL.}
XRPL has no fixed block-size cap; throughput scales inversely with signature size. ML-DSA-44 inflates the signature payload $\sim$38$\times$ (64~B $\to$ 2{,}420~B), Falcon-512 $\sim$10$\times$, and SLH-DSA-128f $\sim$270$\times$, with TPS dropping proportionally.
 
\begin{table}[htbp]
\caption{Theoretical TPS under classical and PQ signature schemes.}
\centering
\footnotesize
\setlength{\tabcolsep}{4pt}
\renewcommand{\arraystretch}{1.1}
\begin{tabular}{|l|c|c|c|c|}
\hline
\textbf{Chain} & \textbf{Classical} & \textbf{ML-DSA-44} & \textbf{Falcon-512} & \textbf{SLH-DSA} \\
\hline
Bitcoin   & 15                      & 1.6           & 3.5            & 0.4   \\
\hline
Ethereum  & 238                     & ${\leq}42$    & ${\leq}105$    & ${\leq}7$  \\
\hline
Avalanche & 476                     & ${\leq}85$    & ${\leq}210$    & ${\leq}14$ \\
\hline
Algorand  & 9{,}295                 & 484           & 1{,}115$^{\dagger}$ & 108   \\
\hline
XRPL      & 3{,}400                 & \multicolumn{3}{c|}{scales as $1/s_{\mathrm{PQ}}$} \\
\hline
Solana    & 65{,}000$^{\star}$      & \multicolumn{3}{c|}{exceeds 1{,}232\,B tx-size cap} \\
\hline
Monero    & 1{,}000                 & \multicolumn{3}{c|}{ring-signature property cannot be replicated$^{\ddagger}$} \\
\hline
\end{tabular}
\\[0.5ex]
\begin{minipage}{\linewidth}
\scriptsize
All values in tx/s. $^{\star}$Protocol-claimed maximum (Solana whitepaper), not derived from the byte-bounded formula. $^{\dagger}$Falcon-512 figure shown for cross-chain comparability; Algorand's deployed scheme is Falcon-1024, which would yield $\approx$585~tx/s native. $^{\ddagger}$A single-signer NIST PQ scheme cannot replace CLSAG without destroying Monero's anonymity (\S~\ref{sec:challenges}); a lattice-based linkable ring signature would be required.
\end{minipage}

\label{tab:blockchain_tps_pq}
\end{table}

\subsection{Aggregation of Consensus Signatures}
\label{sec:agg_loss}

Ethereum's PoS consensus leverages a property unique to BLS signatures:
thousands of individual validator signatures over the same message can be
combined into a single 96-byte aggregate via simple group addition
\cite{boneh2003aggregate}. This is not merely an optimisation---it is what
allows attestations from a network of ${\sim}10^{6}$ validators to be
recorded in only a few KB per slot. With ${\sim}892K$ active validators
on the beacon chain (April~2026 snapshot, a figure declining since Pectra's
increase of the per-validator effective-balance ceiling from 32 to
2{,}048~ETH \cite{beaconchain2026, thedefiant2025decline}), and exactly one
attestation per validator per 32-slot epoch, the network produces roughly
28K attestations per slot, distributed across ${\sim}64$ beacon committees.
Each committee aggregates its members' signatures into a single BLS
aggregate plus a participation bitfield, so the chain stores on the order
of a few KB of attestation data per slot rather than tens of MB.
 
None of the three NIST-standardised PQ signature schemes admits comparable
native aggregation. At 28K attestations per slot, storing unaggregated PQ signatures individually would consume 67.7~MB (ML-DSA-44), 18.6~MB (Falcon-512), or 478~MB (SLH-DSA-128f) per slot---orders of magnitude beyond the current few-KB BLS aggregate 
and the latter two would exceed Ethereum's current per-slot blob and
execution envelope outright. This comparison deliberately sets a real
BLS-aggregated deployment against an unaggregated PQ strawman; no
realistic proposal would deploy any of these schemes in this configuration.
 
The practical mitigation is to aggregate PQ signatures inside a succinct
proof: one SNARK or STARK attests that all underlying signatures verify,
recovering most of the size savings at the cost of verifier overhead,
trusted setup (SNARKs) or larger proofs (STARKs). Three lines of work are
actively pursuing this. (i)~The Ethereum Foundation's PQ roadmap
targets hash-based multi-signatures with STARK-friendly verification as
the eventual replacement for BLS aggregation
\cite{drake2025hashbased,btq_eth_roadmap}.
(ii)~Privacy~\&~Scaling Explorations has demonstrated end-to-end
aggregation of Falcon signatures via LaBRADOR-style lattice arguments
\cite{pse_falcon_aggregation}.
(iii)~Independent prototypes---e.g., BTQ and StarkWare's on-chain Falcon
verifier on StarkNet \cite{btq_falcon_starkware}---show that the verifier side
is tractable in production zk environments, though end-to-end aggregation
throughput at validator-count scale is still open.
 
Aggregation loss is therefore a structurally distinct category of
disruption from raw signature-size growth,
and it can dominate throughput degradation even when individual signatures
appear manageable. We formalise the resulting requirement as~R3 in
\S\ref{sec:pq_requirements}: any blockchain-suitable PQ signature must
support either native aggregation comparable to BLS or an efficient
SNARK/STARK aggregator, since otherwise the consensus-layer payload
becomes the binding constraint long before the transaction-layer payload
does.

\subsection{Public-Key and Storage Footprint}

PQ adoption inflates not only signatures but, for lattice schemes, public keys. Relative to Ed25519's 32-byte PK, ML-DSA-44 is 1,312 bytes ($\sim$41$\times$) and Falcon-512 is 897~bytes ($\sim$28$\times$); SLH-DSA-128f retains a 32-byte PK but offsets this with 17{,}088-byte signatures. This affects ledger storage, per-transaction calldata on account-based chains, and on-chain identity registries.
 
Table~\ref{tab:blockchain_storage_pq} reports \emph{worst-case upper bounds} on storage under ML-DSA-44 substitution, equivalent to assuming signatures and public keys account for 100\% of stored data. The applicable sig+PK growth ratio depends on the chain's underlying classical scheme: for ECDSA chains (Bitcoin, Avalanche, Ethereum tx layer) the ratio is $\sim$36$\times$ (104B$\to$3{,}732~B), while for Ed25519 chains (Algorand, Solana, Monero, XRPL) it is $\sim$39$\times$ (96B$\to$3,732B); we apply a conservative range of $\sim$34--40$\times$ across rows. In practice, block headers, scripts, UTXO commitments, OP\_RETURN data, and per-block metadata are largely unaffected; if sig+PK constitute a fraction $f\in[0.25,\,0.50]$ of total bytes, the effective growth factor falls to $\approx$9--18$\times$, placing realistic ML-DSA-44 storage 2.5--4$\times$ below the tabulated worst-case. Falcon-512 roughly halves the worst-case figures (sig+PK $\approx$1.6~kB vs.\;ML-DSA-44's $\approx$3.7~kB); SLH-DSA-128f roughly doubles them.
 
\begin{table}[htbp]
\caption{Worst-case ledger storage under ML-DSA-44 substitution.}
\centering
\footnotesize
\setlength{\tabcolsep}{4pt}
\renewcommand{\arraystretch}{1.1}
\begin{tabular}{|l|l|l|}
\hline
\textbf{Chain} & \textbf{Full / validator} & \textbf{Pruned / archive} \\
\hline
Bitcoin             & $\sim$21--24 TB        & $\sim$350--400 GB (pruned) \\
\hline
Ethereum (Geth)     & $\sim$22--26 TB        & $\sim$420--480 TB (archive)   \\
\hline
Ethereum (Erigon)   & $\sim$32--37 TB        & $\sim$63--72 TB (archive)  \\
\hline
Avalanche           & $\sim$12--14 TB        & $\sim$70--80 TB (archive)  \\
\hline
Monero$^{\ddagger}$ & ---         & --- \\
\hline
Solana              & 35--160 TB             & $\sim$10--16 PB (archive)  \\
\hline
Algorand            & $\sim$0.7--0.8 TB      & $\sim$105--124 TB (archive)\\
\hline
XRPL                & $\sim$910--1{,}040 TB  & $\sim$1.8--2 TB(non-full-history)                        \\
\hline
\end{tabular}
\\[0.4ex]
\begin{minipage}{\linewidth}
\scriptsize
Worst-case upper bound under ML-DSA-44 substitution, applying a sig+PK growth factor of $\sim$34--40$\times$ (chain-dependent on whether the original scheme is ECDSA or Ed25519) to all on-chain bytes. Realistic figures are 2.5--4$\times$ lower (sig+PK fraction $f \in [0.25, 0.50]$). Falcon-512 roughly halves these bounds; SLH-DSA-128f roughly doubles them. $^{\ddagger}$ML-DSA-44 cannot replace CLSAG ring signatures 
\end{minipage}

\label{tab:blockchain_storage_pq}
\end{table}

\subsection{Block Interval Time}

The block interval is a consensus parameter, not a function of transaction cost: PoWchains (Bitcoin, Monero) hold it near a target through difficulty retargeting, and the remaining chains fix it by schedule. Substituting a PQ scheme therefore does not change the interval itself---it only consumes the \emph{slack within} it. Table~\ref{tab:blockchain_blockintervals} lists the current intervals.

\begin{table}[htbp]
\caption{Current block intervals across the surveyed blockchains.}
\centering
\small
\begin{tabular}{|l|c|}
\hline
\textbf{Blockchain} & \textbf{Block Interval} \\
\hline
Bitcoin             & $\sim$10 min       \\
\hline
Ethereum            & $\sim$12 s         \\
\hline
Avalanche (C-Chain) & $\sim$1.4--1.8 s   \\
\hline
Monero              & $\sim$2 min        \\
\hline
Solana              & $\sim$0.4 s        \\
\hline
Algorand            & $\sim$2.82 s       \\
\hline
XRPL                & $\sim$4 s          \\
\hline
\end{tabular}

\label{tab:blockchain_blockintervals}
\end{table}

Of the two costs that grow, verification is negligible: the NIST schemes verify in well under a millisecond---ML-DSA in about $0.14$~ms, faster than ECDSA at high security levels---so even tens of thousands of checks per block stay far inside any chain's interval~\cite{schemitt2025assessing}. The binding effect is size: a block with $k$ signatures grows by roughly $k\,(|\sigma_{\mathrm{PQ}}|-|\sigma_{\mathrm{classical}}|)$ bytes, and larger blocks propagate more slowly.
 
It is propagation delay, not the interval, that actually moves. Decker and Wattenhofer identify it as the primary cause of Bitcoin forks~\cite{decker2013propagation}, and Gervais et al.\ model the stale (orphan) rate as rising with block size at a fixed interval, narrowing the margin against double-spending and selfish mining~\cite{gervais2016security}. The effect is therefore regime-dependent: under PoW, retargeting keeps the mean interval constant, so the cost surfaces as a higher orphan rate, and holding that rate fixed would require \emph{deliberately} lengthening the interval; in fixed-slot chains the slot cannot move on its own, so the larger budget eats into intra-slot deadlines and raises missed-slot rates, relievable only by a governance change to slot time. In every case the interval stays fixed under PQ substitution; what shrinks is the safety margin---higher fork, orphan, and missed-slot rates---which we analyse in \S~\ref{sec:shifts}.

\section{Emerging Effects of PQ Adoption}
\label{sec:shifts}
Beyond raw performance, PQ adoption induces systemic and behavioral shifts in block processing, consensus dynamics, and miner incentives. We address \textit{RQ4: what side effects and behavioral changes does PQ adoption induce?}

\subsection{Convergence of Consensus}
 
Larger PQ signatures push systems toward larger blocks, and larger blocks take longer to propagate across the network. As propagation delay approaches the block interval, nodes fall out of sync: in PoW chains this raises the fork and orphan rates, and in PoS or PoH chains validators miss slots or rounds.
 
The Bitcoin case is the most directly quantified. Klarman et al.~\cite{klarman2019bloxroute} model the fork probability as $P(\text{fork}) = 1 - e^{-t_{90}/t_B}$, where $t_{90}$ is the time required for a block to reach 90\% of the network and $t_B$ is the mean block interval. With Bitcoin's $t_B = 600$~s and an empirically observed $t_{90} = 11.6$~s (March 2017 measurement), the baseline fork probability is 1.915\%.
The natural antecedent for a Bitcoin PQ-migration scenario follows from \S~\ref{sec:challenges}: an ML-DSA-44 P2WPKH transaction is approximately 1{,}016.5~vB versus 109.5~vB classical ($\sim$9.3$\times$ larger). If block-utilisation patterns are preserved, total block size grows by roughly the same factor, and assuming $t_{90}$ scales linearly with block size we get $t_{90} \approx 107.9$~s and a fork probability of $\approx$16.47\%. Comparing the two regimes---1.92\% versus 16.47\%, an $\sim$8.6$\times$ increase---and extending to a 6-confirmation window (assuming independent forks as a coarse sensitivity), $(P_{\text{fork}})^6$ inflates by a factor of $(0.1647/0.01915)^6 \approx 4.0 \times 10^5$, indicating a qualitative shift toward catastrophic safety degradation. Nakamoto's exact recursive catch-up formula~\cite{nakamoto2008bitcoin} yields different absolute values but the same direction; the independence ratio is informative as a sensitivity bound, not an absolute probability. Falcon-512 substitution gives a smaller $\sim$4.3$\times$ size growth, while SLH-DSA-128f reaches $\sim$40$\times$, dramatically amplifying the fork-rate impact.
 
Equivalent quantitative measurements for the other chains are not available in the literature we surveyed, but the qualitative direction is the same. In Algorand, propagation delay grows linearly with block size~\cite{gilad2017algorand}; PQ-era blocks therefore lengthen consensus rounds and reduce agreement throughput. In Solana, missed slots have been attributed to propagation delays in the Turbine broadcast layer~\cite{validatorsapp_log}; larger PQ blocks amplify the same mechanism. PoS designs that depend on signature aggregation (e.g., Ethereum's BLS attestations) face an additional, distinct effect: when aggregation breaks, per-slot signature payloads do not merely grow with block size---they grow linearly with validator count.
 
Without re-tuning protocol parameters, the slowdown in block dissemination and validation will degrade convergence guarantees across all consensus families.

\subsection{Financial Incentives}
 
Miner block-size choices balance fee revenue against propagation-induced orphaning risk. Rizun~\cite{Rizun2016ATF} models expected miner profit as $(1 - \pi(Q)) \times F(Q)$, where $F(Q)$ is the cumulative fee function increasing with block size $Q$ and $\pi(Q)$ is the orphan probability, which grows with propagation time. Houy~\cite{Houy2014} similarly shows that adding transactions both increases fees and raises orphan probability, leading rational miners to self-limit block size. Chen et al.~\cite{chen2021evolutionaryequilibriumanalysisdecision} model miners' block-size choices as an evolutionary game and find that, although fees rise with block size, larger blocks introduce higher latency and lower acceptance probability. Dimitri~\cite{dimitri} finds that miner revenue is bounded: the marginal delay cost outweighs additional fee gains beyond an optimum.
 
PQ signatures slow propagation, raising orphan probability and reducing miner expected payoff at any given block-size choice. In equilibrium, this lowers profitability and intensifies centralisation pressure toward operators that can offset network latency.

The economic analysis above pertains to PoW chains, where miner block-size choice drives orphan dynamics. PoS chains face structurally different incentives: under PQ-induced propagation delay, validators face missed-slot penalties (LMD-GHOST attestation rewards, RANDAO inclusion, Snowman vote-tally inclusion, RPCA timeout), which scale with the slot length and not with block fees. The behavioral effect is centralization toward operators with the lowest network latency; the economic mechanism is fundamentally different from the PoW orphan-rate dynamic. Adjacent work characterizes geographical/latency-driven validator centralization in classical PoS \cite{cortes2023unveiling, heimbach2025geographical} and the engineering response to PQ signature growth on Ethereum \cite{drake2025hashbased,ethereum2026pq}, but neither line of work models the joint effect of PQ-induced propagation delay on PoS reward distribution.

\section{Requirements for Blockchain Suitable PQ Signatures}
\label{sec:pq_requirements}
 
Synthesising the vulnerability, defense, and performance analyses above, we extract the requirements a PQ signature scheme must satisfy to be a credible replacement for ECDSA/Ed25519/BLS in blockchain settings.
 
\noindent\textbf{(R1) Small signatures and small public keys.} Both contribute multiplicatively to ledger storage and bandwidth (\S~\ref{sec:challenges}). Falcon-512 and the candidates in NIST's Onramp targeting $<$1~kB total sig+PK directly address this; ML-DSA-44 is acceptable but not preferred for size-bound chains; SLH-DSA's 17~kB signatures rule it out as a default for high-throughput L1.
 
\noindent\textbf{(R2) Fast, deterministic verification.} Block validators verify many signatures per block under hard timing budgets; verification-time variance disrupts PoH-style designs with sub-second slots. ML-DSA and Falcon meet this bar; SLH-DSA does not.
 
\noindent\textbf{(R3) Native aggregation, or efficient SNARK/STARK aggregation.} For BLS-dependent designs (\S~\ref{sec:agg_loss}), this is the most consequential single requirement. Lattice schemes with native aggregation are an active research direction; absent such constructions, deployments must invest in PQ-secure SNARK/STARK aggregators.
 
\noindent\textbf{(R4) Threshold-friendliness.} Custody ecosystems demand efficient $t$-of-$n$ threshold variants. Threshold ML-DSA constructions are emerging in recent preprints~\cite{niot2026threshold_mldsa, kao2026fips204_threshold, kao2026talus}. Threshold Falcon is harder due to Gaussian sampling and has seen only limited progress: hash-and-sign masking-friendly variants such as Plover\cite{plover2024} and early threshold treatments like LUOV or Falcon sharing\cite{cozzo2020sharing} explore the design space. Masking-friendly designs like Raccoon \cite{raccoon2024crypto} and its threshold variant \cite{thresholdraccoon2024} sidestep the issue by avoiding Gaussian sampling entirely. SLH-DSA threshold variants are even less developed; the hash-based structure resists conventional MPC techniques, and discussion remains largely informal\cite{bitcoindev2024sphincsthreshold}.
 
\noindent\textbf{(R5) Statelessness.} Hash-based one-time-signature schemes (XMSS, W-OTS) require strict key-state tracking; state loss results in signature forgery. Stateless schemes (ML-DSA, Falcon, SLH-DSA) are safer to deploy at scale; stateful schemes are deployable only in narrow opt-in roles (e.g., the third-party Winternitz Vault program available on Solana, not a network primitive).
 
\noindent\textbf{(R6) Audited library support.} Consensus-critical code cannot be deployed against unvetted reference implementations. Open-source library availability (liboqs, BoringSSL PQ, libpqcrypto) is a prerequisite, not an afterthought.
 
\noindent\textbf{(R7) Crypto-agility-friendly transaction encoding.} The scheme must fit a transaction format whose version negotiation can absorb future replacements without a hard fork (\S~\ref{sec:open_prob}).
 
No single NIST-standardised scheme dominates against all seven requirements. ML-DSA is the strongest default candidate (R2, R4, R5, R6 met); Falcon dominates on (R1) and is competitive on (R2, R5, R6) but lags on (R4) until threshold Falcon matures; SLH-DSA is conservative under (R5, R6) but fails (R1, R2, R3). The Onramp's emphasis on size, performance, and non-lattice diversity~\cite{nist-onramp-2024} indicates that the strongest blockchain-suitable PQ signature has not yet been standardised; chains designing for migration today should prioritise agility (R7) so that future Onramp winners can be adopted without a second hard fork.

\section{Open Problems}
\label{sec:open_prob}
We group the unresolved problems into five categories.

\noindent\textbf{Transition.} Migrating non-PQ accounts to PQ-protected ones must preserve user identity and asset ownership without exposing classical secrets. The core difficulty is proving control of a legacy public key whose corresponding private key is itself the target of the quantum threat. ZK proofs offer a partial path: Baldimtsi et al.~\cite{baldimtsi2025eddsa_migration} re-bind ownership on EdDSA chains using the RFC 8032 seed as a witness, and zk-STARK circuits address the same problem for Bitcoin and Ethereum~\cite{ethereum_btc_zkstark_migration_2026}, though the proof systems are not yet end-to-end quantum-secure under HNDL. The problem is sharpened by large dormant address sets---Satoshi-era Bitcoin outputs, abandoned Ethereum EOAs---whose owners may never migrate proactively. During the migration period, networks can use hybrid signatures to ensure a transaction only verifies if both are valid. This approach hedges against algorithmic failures in either scheme, and both NIST and the NSA explicitly frame hybrid signatures as a transitional tool~\cite{nistir8547, nsa_csa_pq_2024}. They are costly: a hybrid transaction carries both signatures and keys (about $2.5$~KB for ML-DSA-44${+}$ECDSA, $0.74$~KB for Falcon-512${+}$ECDSA, plus the public keys), pays the verification cost of both schemes, and demands explicit version negotiation, and Campbell finds that even under aggressive precompile assumptions hybrids impose order-of-magnitude per-transaction footprint costs and a $60$--$70\%$ throughput loss on permissionless chains~\cite{campbell2025hybrid}. Deployed and proposed designs reflect this staging---Bitcoin's BIP-360 admits either a PQ-only or a hybrid spend witness~\cite{bip360}, Hedera sequences PQ-TLS ahead of PQ signatures~\cite{hedera_pq_2026}, and PQFabric prototypes hybrids in Hyperledger Fabric~\cite{holcomb2020pqfabric}---so the transition is best understood as a multi-year regime in which transaction-format flexibility, custody-layer threshold readiness, and consensus-layer aggregation alternatives mature together, rather than as a single upgrade.

\noindent\textbf{Confidentiality and Privacy.} Blockchain privacy mechanisms---Pedersen commitments, ring signatures, ZK proofs, homomorphic encryption---rely on elliptic-curve assumptions Shor breaks. In UTXO-based systems (Monero, Zcash), Pedersen hiding is unconditional and survives a quantum adversary, but Pedersen binding and CLSAG unforgeability do not. Candidate replacements~\cite{torres2018lrct, matrict, chipmunkring2025, cryptoeprint:2017/766} address subsets of the required properties at varying cost; no construction yet integrates linkable PQ ring signatures, binding-PQ commitments, and PQ-secure range proofs into a coherent stack. Account-based chains face a parallel problem: their confidential-transaction layers rely on ElGamal-style homomorphic encryption~\cite{ElG85} and rollup proofs whose soundness reduces to elliptic-curve assumptions, and any PQ analogue must compose with smart-contract execution under per-block gas budgets.

\noindent\textbf{Consensus Aggregation.} Ethereum's beacon chain uses BLS aggregation to compress an entire committee's attestations into a 96-byte aggregate signature, allowing the tens of thousands of attestations produced per slot across all committees to be recorded in only a few KB; no NIST-standardised PQ signature admits comparable native aggregation, inflating signature payloads to tens of megabytes per slot (Section~\ref{sec:agg_loss}). The candidate workaround---STARK-based aggregation---recovers most of the size but adds verifier cost in a regime where verification time bounds slot length. Native PQ aggregation matching BLS in both size and verification cost remains open.

\noindent\textbf{Communication.} Validator messaging, RPC endpoints, and light-client connections rely on TLS. PQ-TLS handshakes carry larger key material and incur higher overhead; in protocols with millisecond propagation windows (Solana, Avalanche), this directly translates into fork rates and missed slots. Hedera prioritises PQ-TLS as the lowest-risk first step~\cite{hedera_pq_2026} because it requires no consensus changes, but quantifying its impact under the larger payloads imposed by BLS replacement remains an open systems problem.

\noindent\textbf{Custody and Crypto-Agility.} Two ecosystem properties gate practical deployment. First, \emph{threshold-friendly key management.} Exchanges and custodians distribute control of signing keys using threshold and MPC protocols---FROST for Schnorr, GG18/CMP for ECDSA, two-party EdDSA---whose efficiency rests on the linearity of the ECC signing equation in the secret key. Lattice signatures lack this structure: ML-DSA's rejection sampling is non-linear in the secret key, and Falcon's Gaussian tree-sampling resists distribution. Threshold ML-DSA has only recently been demonstrated, in still-unrefereed preprints reporting efficiency up to roughly six parties with sub-second signing~\cite{niot2026threshold_mldsa, kao2026fips204_threshold, kao2026talus}, while threshold Falcon and SLH-DSA remain immature. A protocol-level migration that outpaces its custody ecosystem must therefore either keep ECC signing for the highest-value hot wallets---leaving the largest concentrations of value quantum-vulnerable on a quantum-resistant ledger---or move funds out of threshold custody, re-introducing single-point-of-compromise risk; several projects accordingly treat threshold readiness as a transition prerequisite~\cite{algorand_technology_postquantum, hedera_pq_2026}. Second, \emph{crypto-agility:} the ability to swap schemes without a hard fork as algorithmic failures or new Onramp standards emerge. Recent work explicitly designs for this~\cite{post_quantum_blockchains_agility_2026, fukuda2025grand_challenge}, but---like the hybrid transition above---it requires version-negotiated transaction formats and validator-side multi-scheme support that current chains do not provide by default.
\section{Conclusion}
\label{sec:conclusion}

The emergence of quantum computing poses a profound disruption to blockchain security and performance across contemporary distributed ledger systems. The core reliance on elliptic-curve and hash-based cryptography exposes fundamental vulnerabilities to quantum algorithms, necessitating a transition toward post-quantum alternatives. Integrating quantum-resistant primitives introduces significant performance trade-offs, affecting scalability, latency, and consensus efficiency. In this paper, we study how post-quantum attackers reshape these dynamics, revealing that quantum adaptation extends beyond cryptographic substitution to encompass systemic shifts in protocol design, governance, and interoperability. Ensuring resilience in the quantum era will require coordinated standardization of post-quantum schemes, adoption of hybrid cryptographic infrastructures, and a sustained balance between security, performance, and decentralization.

\bibliographystyle{IEEEtran}
\bibliography{bib/main}

\appendices
\section{Real vs. Theoretical TPS of Blockchains}
\label{sec:real_v_the}

A comparison of TPS in the seven surveyed blockchains is shown in Table \ref{tab:blockchain_tps}.

\begin{table}[!h]
\caption{Comparison of throughput (transactions per second) across different blockchain platforms under classical configurations.}
\centering
\begin{tabular}{|l|c|c|}
\hline
\textbf{Blockchain} & \textbf{Real-time TPS} & \textbf{Theoretical TPS} \\
\hline
Bitcoin & $\sim$ 3--7  & $\sim$ 15 \\
\hline
Ethereum & $\sim$ 15--30  & $\sim 238$  \\
\hline
Avalanche (C-Chain) & $\sim$30--50 & $\sim476$  \\
\hline
Monero & $\sim$ 4 & $\sim 1000$  \\
\hline
Solana & $\sim$ 2,000--4,000  & $\sim$ 65,000$^{\ddagger}$  \\
\hline
Algorand & $\sim$ 10--100 & $\sim$ 9,295 \\
\hline
XRP Ledger (XRPL) & $\sim$ 20--30  & $\sim$ 3,400  \\
\hline
\end{tabular}
\\[0.3ex]
\scriptsize{$^{\ddagger}$Solana figure is the protocol-claimed maximum from the Solana whitepaper, not derived from the byte-bounded formula used for the other chains. Other classical figures use post-Pectra Ethereum (60M gas/12s slot, Nov 2025) and Avalanche C-Chain ACP-176/Octane (15M gas/1.5s slot, April 2025).}

\label{tab:blockchain_tps}
\end{table}

\section{Storage in Current Blockchains}

The current storage requirements of the seven surveyed blockchains is shown in Table \ref{tab:blockchain_storage}.

\begin{table}[!h]
\caption{Approximate storage requirements for different blockchain platforms under classical configurations.}
    \centering
    \begin{tabular}{|l|p{5cm}|}
        \hline
        \textbf{Blockchain} & \textbf{Storage} \\
        \hline

        Bitcoin
        & Full: $\sim$750 GB \\
        & Pruned: $\sim$7 GB \\
        \hline

        Ethereum (Geth)
        & Full: $\sim$1.2 TB \\
        & Archive (hash-based, legacy): $>$20 TB \\
        & Archive (PBSS, v1.16+): $\sim$2 TB \\
        \hline

        Ethereum (Erigon)
        & Full: $\sim$920 GB \\
        & Archive: $\sim$1.8 TB \\
        \hline

        Avalanche (C-Chain)
        & Full: $\sim$1 TB \\
        & Archive: $\sim$2 TB \\
        \hline

        Monero
        & Full: $\sim$245~GiB \\
        & Pruned: $\sim$100~GiB \\
        \hline

        Solana
        & Validator: 2--4 TB \\
        & Archive: $>$400 TB \\
        \hline

        Algorand
        & Full: $\sim$20 GB \\
        & Archive: ColdDataDir $\sim$3 TB + HotDataDir $\sim$100 GB \\
        \hline

        XRP Ledger (XRPL)
        & Full history: $\sim$39 TB \\
        & Non-full-history: $\geq$100 GB \\
        \hline
    \end{tabular}
    \label{tab:blockchain_storage}
\end{table}

\end{document}